# A REVIEW OF SECURITY ATTACKS AND INTRUSION DETECTION SCHEMES IN WIRELESS SENSOR NETWORK

Yassine MALEH[1] and Abdellah Ezzati[2]

Emerging Technologies Laboratory (VETE), Faculty of Sciences and Technology
Hassan 1st University, Settat, MOROCCO
[1]`Yassine.maleh@gmail.com`,[2]`abdezzati@gmail.com`

## ABSTRACT

*Wireless sensor networks are currently the greatest innovation in the field of telecommunications. WSNs have a wide range of potential applications, including security and surveillance, control, actuation and maintenance of complex systems and fine-grain monitoring of indoor and outdoor environments. However security is one of the major aspects of Wireless sensor networks due to the resource limitations of sensor nodes. Those networks are facing several threats that affect their functioning and their life. In this paper we present security attacks in wireless sensor networks, and we focus on comparison and analysis of recent Intrusion Detection schemes in WSNs.*

## 1. INTRODUCTION

Recent advances in wireless and micro electronic communications have enabled the development of a new type of wireless network called wireless sensor network (WSN).Wireless sensor networks are associated with vulnerable characteristics such as open-air transmission and self-organizing without a fixed infrastructure [1]. Consequently security of wireless sensor networks (WSN) is the most challenge for this type of network [2]. Intrusion Detection Systems (IDSs) can play an important role in detecting and preventing security attacks. This paper presents a review of the security attacks in wireless sensor network and analyzed some of the existing IDS models and architectures. Finally a comparative study and a discussion of IDS models will be presented.

## 2. RELATED WORK

Wireless sensor networks are not immune to the risks of destruction and decommissioning. Some of these risks are identical to those in Ad-Hoc networks, and others are specific to the sensors. Several articles [6][7][8][9][10] have presented security attacks and issues in WSNs. Intrusion detection system (IDS) defined as the second line of defense after cryptography, allows the detection and prevention of internal and external attacks.
In [18, it is presented a Rule-based IDS called also Signature-based. Most of the techniques in these schemes follow three main phases: data acquisition phase, rule application phase and intrusion detection phase. In [19], it is proposed two approaches to improve the security of clusters for sensor networks using IDS. The first approach uses a model-based on authentication, and the second scheme is called Energy-Saving. IN [21] a hybrid intrusion detection system (HIDS) model has been anticipated for wireless sensor networks. This paper does not promote a solution. Rather, it is a comparative study of existing model of intrusion detection in wireless sensor networks. Our aim is to provide a better understanding of the current research issues in this field.

## 3. SECURITY GOALS IN WSN

We can classify the security goals into two goals: main and secondary. The main goals include security objectives that should be available in any system (confidentiality, availability, integrity and authentication). The other category includes secondary goals (self-organization, secure localization, Time synchronization and Resilience to attacks) [3] [4].

- Confidentiality (Forbid access to unwanted third parties)
- Authentication (Identity verification and validation)
- Availability (Service has to be always available)
- Integrity (Data is exchanged without malicious alteration)
- Self Organization(Every sensor node needs to be independent and flexible enough to be self-organizing and self-healing)
- Secure localization (Sensor network often needs location information accurately and automatically)
- Time synchronization (Sensor radio may be turned off periodically in order to conserve power)
- Resilience to attacks (The covenant of a single node must not violate the security of the whole network). Figure1 below summarizes security goals for wireless sensor network.

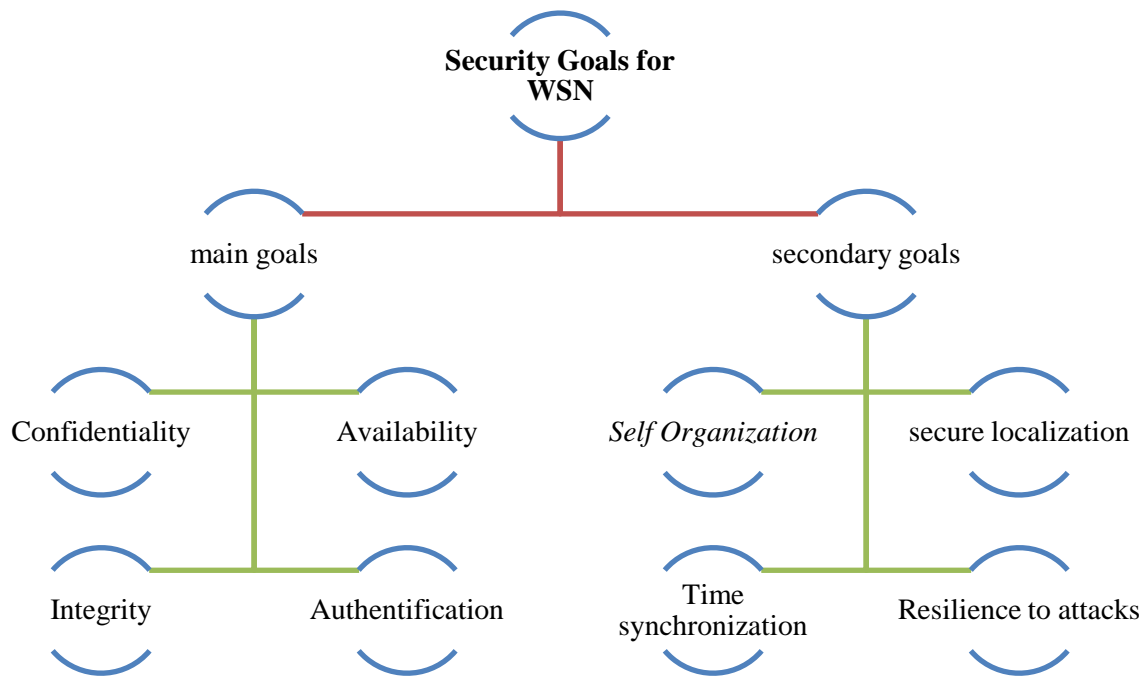

Figure 1. Security Goals for WSN

## 4. SECURITY ATTACKS IN WSN

The different characteristics of wireless sensor networks (energy limited, low-power computing, use of radio waves, etc...) expose them to many security threats. We can classify the attacks into two main categories [5]: Active and Passive. In passive attacks, attackers are typically camouflaged, i.e. hidden, and tap the communication lines to collect data. In active attacks, malicious acts are carried out not only against data confidentiality but also data integrity. Several papers have presented the security attacks in WSN [6][7][8][9][10].

▪ **Spoofed, altered or replayed routing information**

May be used for loop construction, attracting or repelling traffic, extend or shorten source route.

▪ **Selective forwarding**

In this attack, the attacker prevents the transmission of some packets. They will be removed later by the malicious node.

▪ **Worm hole attack:**

The wormhole attack requires insertion of at least two malicious nodes. These two nodes are interconnected by a powerful connection for example a wired link. The malicious node receives packets in one section of the network and sends them to another section of the network.

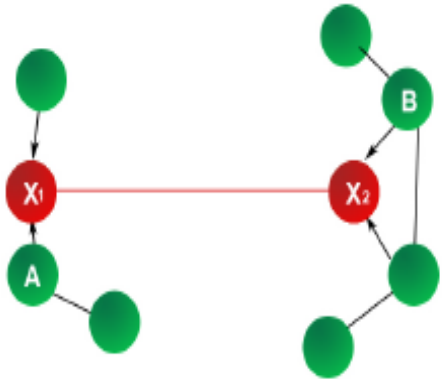

Figure 2. Worm hole attack

▪ **Sybil attack:**

A malicious node presents multiple identities to the other nodes in the network. This poses a significant threat to routing protocols and will cause the saturation of the routing tables of the nodes with incorrect information.

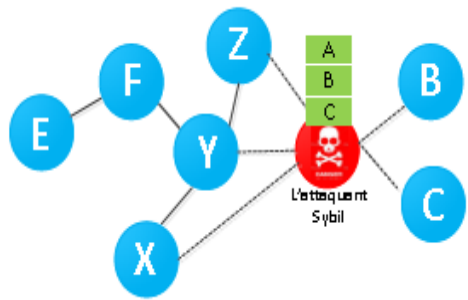

Figure 3. Sybil attack

▪ **Black hole attack:**

The attack involves inserting a malicious node in the network. This node, by various means, will modify the routing tables to force the maximum neighboring nodes passing the information through him. Then like a black hole in space, all the information that will go in it will never be retransmitted.

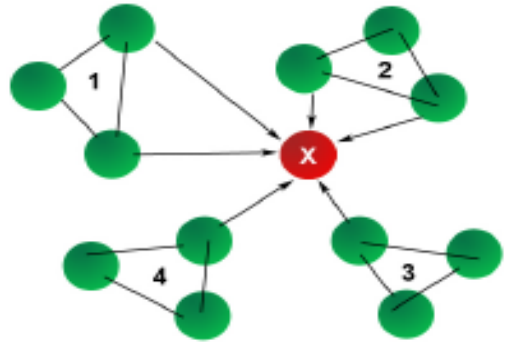

Figure 4. Black hole attack

- **Hello Flooding:**

Discovery protocols on WSNs use HELLO messages types to discover its neighboring nodes. In an attack type HELLO Flooding, an attacker will use this mechanism to saturate the network and consume energy.

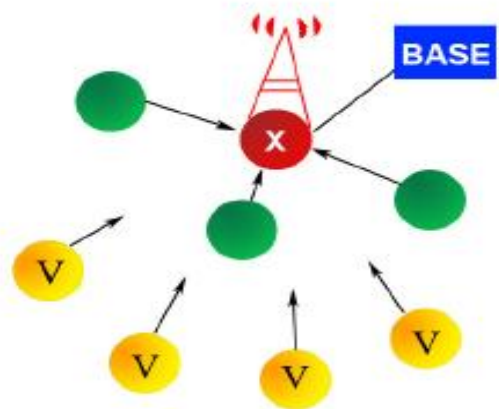

Figure 5. Hello flooding attack

- **Acknowledgement spoofing**

In this attack, the attacker tries to convince the sender that the weak link is strong or that a dead node is alive. Therefore, all packets passing through this link or this node will be lost.

- **Denial-of-Service Attacks**

A denial-of-service (DoS) targets the availability and capacity reduction of network services. Physical constraints of the sensor networks and the nature of their deployment environment, make them vulnerable to DoS attacks more than any other type of network. In this section we will review important DoS scenarios for each layer of the WSN. In [11] Wang *et al.* (2006) have classified the DoS attacks that could target each layer of the WSN.

| Layer | Attacks | Defense |
|---|---|---|
| **Physical** | Jamming | Spread-spectrum, priority messages, lower duty cycle, region mapping, mode change |
| **Link** | Collision | Error-correction code |
| | Exhaustion | Rate limitation |
| | Unfairness | Small frames |
| **Network** | Spoofed routing information, and selective forwarding | Egress filtering, authentication, monitoring |
| | Sinkhole | Redundancy checking |
| | Sybil | Authentication, monitoring, redundancy |
| | Wormhole | Authentication, probing |
| | Hello Flood | Authentication |
| **Transport** | Session Hijacking. | aggregation data |
| | SYN flooding | Package authentication |
| **Application** | Data Corruption. Repudiation | Authentification |

Table 1. Various DOS attacks on WSNs and their countermeasures

## 5. COUNTERMEASURES

To counter the attacks threatened networks wireless sensors, several research teams are trying to find appropriate solutions. These solutions must take into account the specificities of wireless sensor networks. We need to find simple solutions to secure the network while consuming the least possible energy and adapt these solutions to a low power computing. In the range of these solutions include mechanisms such as data partitioning, the use of appropriate cryptographic methods, intruder detection by location or even the confidence index. Wood and Stankovic [12] studied DoS attacks and possible defense. In [13][14] a suite of optimized security protocols for wireless sensor network is presented. SPIN (Security Protocol for Information via Negotiation) has two security mechanisms: SNEP and TESLA. SNEP provides data confidentiality and data authentication. TESLA provides source authentication in multicast scenarios by using MAC chaining.  It is based on loose time synchronization between the sender and the receivers. INSENS (Intrusion Tolerant routing for wireless sensor networks) this protocol allows the base station to draw an accurate map of the network that will establish the routing tables for each node [15]. Du,et al. [16] propose LEAP+ (Localized Encryption and Authentication Protocol), a key management protocol for sensor networks.

## 6. INTRUSION DETECTION SYSTEMS IN WSN

After the concept of intrusion detection (ID), which was established in 1980, two major variants of intrusion detection systems (IDS) have emerged, Host intrusion detection systems (HIDS) and network intrusion detection systems (NIDS) [17]. Intrusion detection is an approach that is complementary with respect to mainstream of security mechanisms such as cryptography and access control [18]. Intrusion detection can be defined as Intrusion detection can be defined as the automatic detection and alarm generation to report that an intrusion has occurred or is in progress. In this section we describe the architecture of IDS in WSNs. IDS cannot take preventive action, since they are passive in nature, they can only detect intrusion and generate an alarm. The following figure presents the four main components of IDS [19].

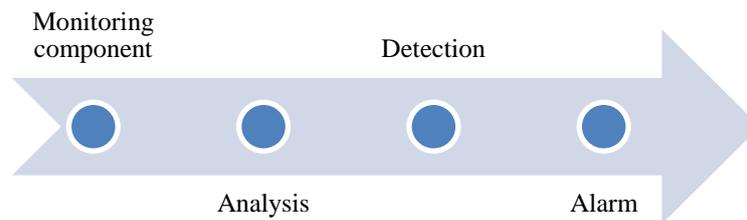

Figure 6. IDS components

There are two distinct technologies of IDS:
- Network Intrusion Detection System (NIDS). These systems are designed to intercept and analyze packets circulating in the network. All communication in the wireless network are conducted on the air and a node can hear the traffic passing from a neighboring node (promiscuous mode) [36]. Therefore, the nodes can mutually check the network traffic. This technology applies this concept, IDS listens for traffic and individually examine each packet.
- Host intrusion detection systems (HIDS). Analysis only data on the node where the IDS is installed. Any decision is based on information collected at this node. These IDSs use two types of sources to provide information about the activity: the log files (file that records all activity on a system in standby), and audit trails ( Incoming / outgoing packets node , etc).

### 5.1 The challenging of designing IDS for WSN

The IDS solutions developed for wired networks cannot be applied directly to sensor networks, view the difference between these two types of networks, this is why it is necessary to introduce an intrusion detection system that meets the special features of sensor networks [20]. The design of this kind of system for wireless sensor network must satisfy the following properties:

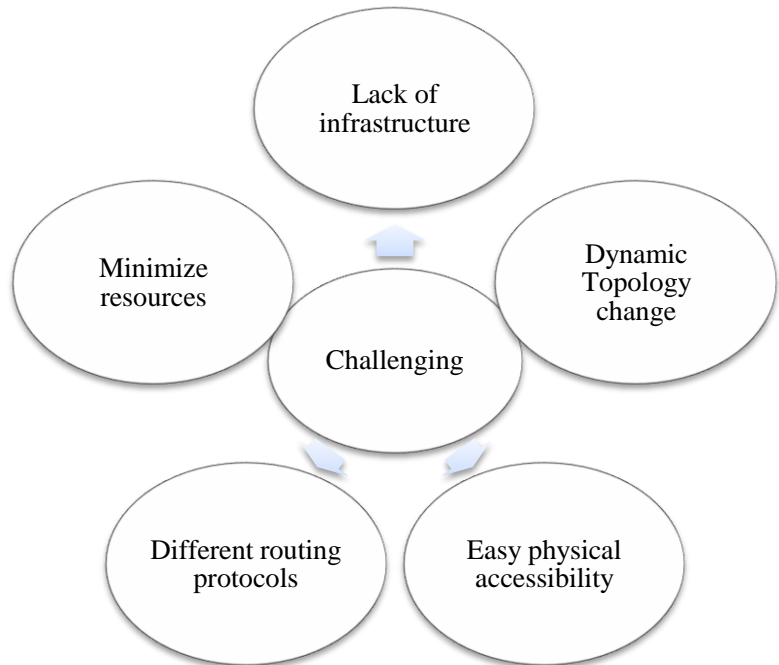

Figure 7. Challenging of designing IDS for WSN

## 5.2 The requirements of designing IDS for WSN

In wireless sensor networks, the IDS must satisfy the following properties [21]:
- Localize auditing: IDS for wireless sensor networks must work with local data and partial audits, because in WSN there are no centralized points (apart from the station base) that can collect global data auditing.
- Minimize resources: IDS must use a minimum number of resources for networks. Communication between nodes for intrusion detection should not saturate the available bandwidth.
- Trust no node: Unlike wired networks, nodes sensors can be compromised easily, IDS must not trust any node.
- Be distributed: means that the collection and analysis of data should be in several locations. Moreover the distributed approach also applies to the execution of the algorithm of detection and alert correlation.
- Be secure: IDS must be able to withstand attacks.

Figure 8 below summarizes requirements of designing IDS for WSN.

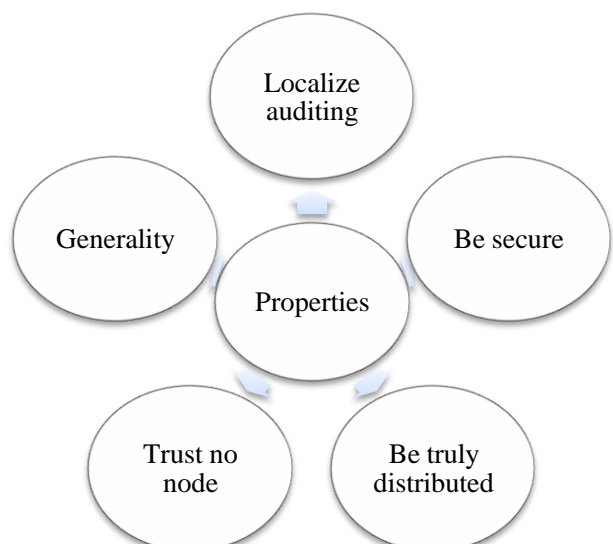

Figure 8. Requirements of designing IDS for WSN

## 5.3 Architectures for IDS in Wireless Sensor Network

The nature of wireless sensor networks makes them very vulnerable to attack. The Mobile nodes are randomly distributed, there are no physical obstacles for the adversary, therefore, they can be easily captured, and attacks can come from all directions and target any node. To tackle these additional challenges, several possible IDS architectures exist including standalone IDS, distributed and cooperative IDS and hierarchical IDS [22].

### 5.3.1 Standalone IDS

In this category, each node operates as independent IDS and is responsible for the detection of attacks against him. Therefore, the IDS do not cooperate and do not share information with each other. This architecture requires that each node is capable of executing and running IDS.

### 5.3.2 Distributed and Cooperative IDS

In this architecture (Zhang et al., 2003), each node has an IDS agent and makes local detection decisions by itself, all the nodes cooperate to create a global detection process. The distributed and cooperative IDS architecture is more suitable for a flat network configuration than a cluster-based multilayered one.

### 5.3.3 Hierarchical IDS

In this category the network is divided into clusters with cluster-heads. In each cluster, a leader plays the role of cluster-head. This node is responsible for routing in the group and must accept messages from members of the cluster indicating something malicious. Similarly, the cluster-head must detect attacks against other cluster-heads in the network. At the same time all cluster-heads can cooperate with central base station to form global IDS.

## 5.4 Some open research in IDS

**Cross-Layer IDS:** Using a cross layer IDS, we could not only pass information between layers but also coordinate mechanisms to prevent threats at all layers.
**Dynamic IDS:** The IDS that would protect mobile nodes, as in VANET networks.
**Internet of Things IDS:** There should be mechanisms that could manage all the objects of our everyday life that have an IP address and be connected to the Internet.

## 7. INTRUSION DETECTION MODELS FOR WSN

Due to architectural difference between wired and wireless networks, their IDSs cannot be used interchangeably. There are specific techniques for WSN [23]. In this section, we analyze and discus some proposed IDSs for WSN.

### 7.1 Rule-based IDS

Rule-based IDS called also Signature-based IDS, articulates on a database of stored prior rules of security attacks [24]. Most of the techniques in these schemes follow three main phases: data acquisition phase, rule application phase and intrusion detection phase (Silva et al., 2005) [25]. The algorithm includes three steps for detecting intrusions. In the first step monitor nodes monitors the data. In the second step detection rules will be ranked in order of severity, to the collected information to flag failure. The third step is the intrusion detection phase, where the number of failure flagged is compared to the expected number of the occasional failures in the network.

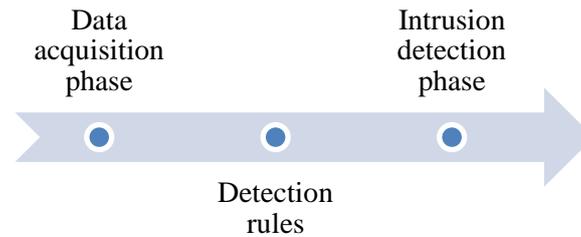

**Figure 9.** Steps for detecting intrusion in rule based IDS

### 7.2 Cluster-Based IDS

Su, et al. [26] has proposed two approaches to improve the security of clusters for sensor networks using IDS. The first approach uses a model-based on authentication, which can resist to external attacks. Its basic technique is to add a message authentication code (MAC) for each message. Whenever a node wants to send a message, it adds to it a timestamp and a MAC is generated by a key-pair or individually depending on the key role of the sender (cluster-head, member -node, or base station). So that the receiver can verify the sender, the security mechanism is used LEAP. The second scheme is called Energy-Saving. This approach focuses on the detection of misbehavior both in Member nodes (MN) and cluster-head nodes (CH). When misbehavior is detected, the CH broadcasts a warning message encrypted with the cluster key to restrain this specific node.

### 7.3 Hybrid IDS

In the Hybrid Approach, both techniques (Cluster-Based and Rule-Based) are combined to form Hybrid detection technique. Hybrid detection exploits the advantages of both approaches provides simplicity, high safety, low consumption of energy [27] [28].The Hybrid Intrusion Detection System achieves the goals of high detection rate and low false positive rate.

### 8. Analyses and Discussion

Comparing analysis, for the advantages and drawbacks of different models:

Rule-Based : The rule based model is simple, clear levels, and designed for a large-sized WSNs. Signature-based IDS need more resource than anomaly-based IDS, and regular updating of the database with new attack signatures.

Cluster-based: The cluster-based model requires Low Energy Consumption, provides high level of security. Because of Centralized routing data delivery is guaranteed. In cluster-based IDS Message retransmission frequency is high, and the centralized routing may not always use best available path for routing.

Hybrid model: Hybrid model are designed for large and sustainable WSN. This model uses two mechanisms, anomaly-based and signature-based, so it requires high consumption of energy. Table 2 gives the comparison and characteristics of different IDSs.

| IDS Model | Network architecture | Detection technique | Handled attacks | Energy consumption | Advantages | Drawbacks |
|---|---|---|---|---|---|---|
| **Anomaly based IDS** | | Anomaly Based | Masquerade, routing attacks, Sinkhole and blackhole | Low | Capable of detecting new attacks | Misses well known attack |
| **Rule-based IDS** | Distributed | Signature based | Black hole, selective forwarding, Sink hole, DOS | Low | Detects all those attacks having signatures | Cannot detect new attacks |
| **Cluster-based IDS** | Hierarchical | Anomaly Based | | Low | Low Energy Consumption  Data delivery is guaranteed | Message retransmission frequency is high, Increased Traffic |
| **Hybrid IDS** | Hierarchical | Anomaly based | Selective forwarding, sinkhole, Hello flood and wormhole attacks | Medium | Can detect both existing and new attacks | Requires more computation and resources |
| **Intrusion detection in Routing Attacks** | Distributed | Anomaly based | DoS, Sinkhole and wormhole attacks | High | Consider resource Parameters (energy and reliability) | High resource Requirement, Increased Traffic |

Table2: Comparison and characteristics of different IDSs model

## 9. Conclusion

This article shows how well a security sensor networks is a challenge for researchers and developers of information technology. Our goal was to present the existing security attacks in WSN, focusing on intrusion detection systems (IDS), and examine existing approaches of intrusion detection in WSN. Our goal was to present the existing security mechanisms for WSN, specifically focusing on intrusion detection systems (IDS), and consider existing approaches to provide a fairly comprehensive and effective model. We are now working on our own model that incorporates all the advantages of the approaches proposed for a global model of intrusion detection in WSN.

**Authors**

**Yassine MALEH** received the B.Sc. degree in networks and Information Systems, from Hassan 1st University, Faculty of Sciences and Technology of Settat, Morocco, in 2009, and M.Sc. degree in Network and Computer engineering from the Hassan 1st University, Faculty of Sciences and Techniques (FSTS), Settat, Morocco, in 2012. Currently pursuing his PhD in Networks and Security Engineering at the Laboratory of Emerging Technologies (VETE), from Hassan 1st University, Faculty of Sciences and Technology of Settat, Morocco. His main research areas are how to use wireless sensor networks to secure and monitor mobile laboratories networks.

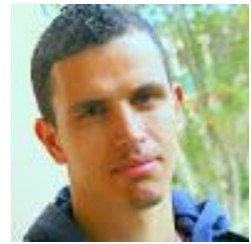

**Abdellah EZZATI** research Scientist in Faculty of Science and Technology in Morocco. He obtained his PHD in 1997 in Faculty of science in Rabat and member of the Computer commission in the same Faculty. Now is an associate professor in Hassan First University in Morocco and he is the Head of Bachelor of Computer Science. He participate to several project as the project Palmes which elaborate a Moroccan Education Certification.

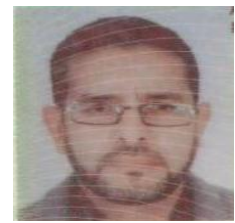